\documentclass{article}

    \usepackage[preprint]{neurips_2025}

\usepackage[utf8]{inputenc} 
\usepackage[T1]{fontenc}    
\usepackage{hyperref}       
\usepackage{url}            
\usepackage{longtable}  
\usepackage{booktabs}       
\usepackage{tabularx}
\usepackage{amsfonts}       
\usepackage{amsmath}
\usepackage{nicefrac}       
\usepackage{microtype}      
\usepackage{xcolor}         
\usepackage{graphicx} 
\usepackage{multirow}
\usepackage{enumitem}
\setlist[itemize]{
    leftmargin=17pt
}

\title{Generative Representational Learning of Foundation Models for Recommendation}

\author{%
  Zheli Zhou$^1$, Chenxu Zhu$^2$, Jianghao Lin$^1$\thanks{Co-corresponding authors.}\;\; \\ \textbf{Bo Chen$^2$, Ruiming Tang$^2$, Yong Yu$^1$, Weinan Zhang$^1$\footnotemark[1]}\\
  $^1$Shanghai Jiao Tong University, $^2$Huawei Noah's Ark Lab \\\texttt{\{zlz114514,chiangel,wnzhang\}@sjtu.edu.cn} \\
  \url{https://junkfood436.github.io/RecFound/}
}

\begin{document}

\maketitle

\begin{abstract}

Developing a single foundation model with the capability to excel across diverse tasks has been a long-standing objective in the field of artificial intelligence. As the wave of general-purpose foundation models sweeps across various domains, their influence has significantly extended to the field of recommendation systems. While recent efforts have explored recommendation foundation models for various generative tasks, they often overlook crucial embedding tasks and struggle with the complexities of multi-task learning, including knowledge sharing \& conflict resolution, and convergence speed inconsistencies. To address these limitations, we introduce \textbf{RecFound}, a generative representational learning framework for recommendation foundation models. We construct the first comprehensive dataset for recommendation foundation models covering both generative and embedding tasks across diverse scenarios. Based on this dataset, we propose a novel multi-task training scheme featuring a Task-wise Mixture of Low-rank Experts (TMoLE) to handle knowledge sharing \& conflict, a Step-wise Convergence-oriented Sample Scheduler (S2Sched) to address inconsistent convergence, and a Model Merge module to balance the performance across tasks. 
Experiments demonstrate that RecFound achieves state-of-the-art performance across various recommendation tasks, outperforming existing baselines.

\end{abstract}

\section{Introduction}


It has been a long-standing goal to develop a single foundation model to excel across diverse tasks in the field of artificial intelligence~\cite{bommasani2021opportunities}. 
Foundation models, characterized by their extensive knowledge bases and large-scale parameters, have demonstrated remarkable generalization capabilities across various domains, including natural language processing and computer vision. These models rely on large-scale pretraining to capture rich representations, enabling them to adapt to a wide range of downstream tasks with minimal finetuning~\cite{muennighoff2025grit}. Recently, the influence of foundation models has extended to the field of recommendation systems, where they offer unique opportunities to enhance personalization and prediction accuracy by effectively modeling complex user-item interactions and contextual information of distinct sources and formats~\cite{huang2024foundation,lin2024rella,lin2024clickprompt,xi2024memocrs,wang2024flip}. 

Early efforts in applying foundation models to recommendation systems primarily focus on specific, relatively simple tasks such as tagging~\cite{konjengbam2020unsupervised} and question answering~\cite{lai2019supervised}. These approaches often involve adapting general-purpose language models to handle individual recommendation-related tasks, demonstrating the potential of foundation models in this domain. 
More recent work has aimed to generalize foundation models to handle a broader spectrum of recommendation tasks~\cite{shi2024llamae, li2024EcomGPT, peng2024eCeLLM, zhang2024LLaSA}. While these works demonstrate promising progress in unifying different recommendation tasks under a single model, they still face the following significant challenges in effectively handling the full complexity of recommendation scenarios. 

\textit{First, existing approaches overlook the crucial embedding\footnote{In this work, the term ``embedding'' is interchangeable with ``representational''.} tasks}, which are fundamental to recommendation performance. Embedding techniques transform high-dimensional, discrete features such as user and item identifiers or descriptions into low-dimensional, continuous vectors, capturing intricate relationships and facilitating efficient computation. 
High-quality embeddings are essential for various components of recommendation systems, including retrieval, recall, and visualization~\cite{huang2024comprehensive,liu2023embedding, chen2022deep}. 
\textit{Second, current models struggle with the complexities of multi-task learning}, particularly in managing the knowledge sharing \& conflict and the inconsistent convergence speeds among tasks. These issues can hinder the effective training of foundation models intended to perform well across diverse recommendation tasks, limiting their practical applicability in real-world systems~\cite{xi2024towards,wu2023robust}.

To address these limitations, we introduce \textbf{RecFound}, a generative representational learning framework for recommendation foundation models. Our framework begins with the construction of the first comprehensive dataset that covers both generative and embedding tasks across various aspects of the recommendation domain. Building upon this dataset, we propose a novel multi-task training scheme for recommendation foundation models. 
From the model perspective, we design a \textbf{Task-wise Mixture of low-rank experts (TMoLE)} architecture. TMoLE consists of multiple low-rank experts and a task‑oriented router. 
The router generates a masked routing distribution based on each sample’s task to adaptively aggregate the expert outputs. 
This mechanism allows for effective knowledge sharing between related tasks while mitigating potential conflict among disparate ones by enabling task-specific specialization via the expert integration. 
From the data perspective, we propose a \textbf{Step-wise convergence-oriented sample scheduler (S2Sched)}. 
S2Sched dynamically adjusts the sampling ratio of different tasks within each training batch based on their non-convergence rates estimated on the validation set. 
Specifically, it prioritizes sampling from tasks that are converging more slowly, thereby addressing the issue of inconsistent convergence speeds across different recommendation tasks within a single model.
Finally, we also introduce a \textbf{Model Merge} module to further balance the perfromance across tasks.

Our contributions can be summarized as follows:
\begin{itemize}
    \item We construct the first comprehensive dataset for recommendation foundation models, covering both 10 generative tasks and 3 embedding tasks across diverse recommendation scenarios.
    \item We propose RecFound, a novel training framework for recommendation foundation models. We design TMoLE to facilitate effective knowledge sharing \& conflict resolution between embedding and generative tasks, S2Sched to address the inconsistency of convergence speeds for different tasks, and the Model Merge module to balance performance across tasks.
    \item Extensive experiments demonstrate that our RecFound framework achieves state-of-the-art performance on a wide range of recommendation tasks, and we provide insightful analysis to support our findings.
    We release our code, as well as the dataset and model checkpoints, to benefit the research community.
\end{itemize}








\section{Related Works}

\paragraph{General Foundation Models in NLP}
Extensive research on general foundation models in natural language processing (NLP) has investigated general-purpose methods for addressing a wide range of downstream tasks.
To solve various embedding tasks, early influential works focused on learning static word embeddings from large corpora~\cite{mikolov2013word2vec,pennington2014glove}. 
Subsequent research, such as ELMo~\cite{peters2018elmo}, BERT~\cite{devlin2019bert}, and a wave of BERT-family models~\cite{liu2019roberta, reimers2019sentence} generates token representations that dynamically reflect their surrounding context, yielding substantial gains in tasks like semantic matching and retrieval.
Another line of work investigates foundation models designed for various generative tasks, aiming to solve multiple types of text generation tasks simultaneously. 
The GPT series~\cite{radford2018gpt1,radford2019gpt2,brown2020gpt3} demonstrated this by revealing how decoder-only architectures can acquire multitask proficiency only through next-token prediction.
More recently, a few studies have explored the possibility of a unified foundation model that can effectively handle both embedding and generative tasks. 
GritLM~\cite{muennighoff2025grit} trains a large language model to perform both embedding and generative tasks by distinguishing them through instructions. 
Similarly, UniGen~\cite{li2023unigen} integrates retrieval and question answering into a single model.

\paragraph{Recommendation Foundation Models}
Early research on recommendation focused on optimizing single-task generative models~\cite{hidasi2015gru4rec,dong2017seq2seq,wang2017irgan,liang2018variational,lin2025can,xi2024decoding}, such as producing personalized reviews via neural review generators~\cite{dong2017neural_review_gen} or generating item descriptions from user behavior sequences~\cite{xi2024efficient,yang2018behavior_driven_desc}. These models are typically trained from scratch and lack generalization across tasks. In the wake of large language models, recent works introduce foundation models tailored to e-commerce scenarios~\cite{shen2023large,zhou2024large,wang2024unirec,liu2023fim}. LLaMA‑E \cite{shi2024llamae} tunes LLaMA~\cite{grattafiori2024llama3herdmodels} to support various product authoring tasks. EcomGPT~\cite{li2024EcomGPT} leverages generative pretraining over large-scale e-commerce corpora to unify tasks such as product Q\&A, summarization, and marketing text generation. 
eCeLLM~\cite{peng2024eCeLLM} further develops a general-purpose e-commerce model by training on high-quality instruction datasets, enabling both in-domain and cross-task generalization. Such efforts demonstrate the growing interest in generative foundation models for recommendation. However, the field still lacks foundation models that support both representation and generative capabilities.
To the best of our knowledge, we are the first to develop a unified generative representational learning framework for recommendation foundation models.

\section{RecFound Dataset}

We present \textbf{RecFound}, a meticulously curated dataset that unifies embedding (i.e., representational) and generative tasks within the recommendation domain. Our dataset possesses three key advantages.
(1) \textbf{Innovation}: We are the first to integrate embedding and generative tasks in recommendation systems, providing a novel benchmark and perspective for research;
(2) \textbf{Diversity}: The dataset covers multiple recommendation tasks and domains, including e-commerce, social media, music, video streaming, and news recommendations. Each data sample has rich expressiveness to facilitate model training and evaluation;
(3) \textbf{High-Quality}: All data originates from real user interactions, with rigorous cleaning to remove samples containing illegal symbols, errors, or missing content.

We give a brief introduction to each embedding and generative task below, with detailed data examples provided in Appendix~\ref{app:data}.

\textbf{Embedding Tasks} include three classical representational learning scenarios for recommendation:
\begin{itemize}
      \item \textbf{User2Item (U2I)}: U2I task involves encoding a user's behavior sequence into an embedding that encapsulates their interests. The output user representation is then utilized to retrieve relevant items. The samples for this task are constructed on Amazon Reviews dataset~\cite{xu-etal-2019-scaling,he2016ups,mcauley2015image,hou2024bridging}.

      \item \textbf{Query2Item (Q2I)}: In Q2I task, user input queries are embedded to provide precise item recommendations based on the query embeddings. This task is built upon Shopping Queries~\cite{reddy2022shopping} and ECInstruct datasets~\cite{peng2024eCeLLM}.
    
      \item \textbf{Item2Item (I2I)}: I2I task focuses on recommending items similar to a given target item by encoding the embeddings of different item descriptions. All data for this task is derived from Amazon Reviews dataset~\cite{xu-etal-2019-scaling,he2016ups,mcauley2015image,hou2024bridging}.
\end{itemize}

\textbf{Generative Tasks} consist of 10 subtasks, which can be categorized into three different types.
\begin{itemize}
    \item \textbf{General NLP}. 
    We use MuDoCo~\cite{martin-etal-2020-mudoco}, Amazon Reviews~\cite{xu-etal-2019-scaling,he2016ups,mcauley2015image,hou2024bridging}, and AmazonQA datasets~\cite{gupta2019amazonqa}, and construct three tasks to provide the model with better language capabilities to solve recommendation questions: 
    (i) Query Rewriting (QR) aims to rewrite obscure queries from users for a better understanding.
    (ii) Attribute Value Extraction (AVE) expects to extract attribute values of a given item, including titles, descriptions, brands, etc. 
    (iii) Answer Generation (AG) is the common Q\&A task with question topics related to recommendation scenarios. 
    
    \item \textbf{User Understanding}: 
    We adopt Amazon Reviews~\cite{xu-etal-2019-scaling,he2016ups,mcauley2015image,hou2024bridging} and MovieLens datasets~\cite{10.1145/2827872}, and develop four tasks for a better understanding of users' behaviors, interests, emotions, and intentions. 
    (i) Sequential Recommendation (SR) aims to predict what the user will purchase next given his/her history behavior sequence. 
    (ii) Sentiment Analysis (SA) is a task to estimate the emotion and preference based on users' textual reviews.
    (iii) User Profile (UP) requires the model to reason and summarize users' detailed preferences based on his/her behavior history.
    (iv) Answerability Prediction (AP) 
    judges whether the questions from users can be answered or not. 
    
    \item \textbf{Item Understanding}:
    We absorb Amazon Reviews ~\cite{xu-etal-2019-scaling,he2016ups,mcauley2015image,hou2024bridging}, Amazon-Google Products~\cite{kopcke2010evaluation, Rahm2010EM}, and MovieLens~\cite{10.1145/2827872} datasets to build three tasks aiming at extracting effective information from items. 
    (i) Product Relation Prediction (PRP) conducts three-tier relevance prediction for a given item pair.
    (ii) Product Matching (PM) judges whether the given two items are the same when they contain different descriptions from different platforms.
    (iii) Item Profile (IP) requires the model to extract crucial factual information of a given item.
\end{itemize}

We merge the data samples from all the tasks to form our RecFound dataset, which consists of 3 embedding tasks and 10 generative tasks.
Next, we split the dataset into training, validation, and testing sets. We report the detailed dataset statistics in Table~\ref{tab:data}.

\begin{table}[t]
\centering
\caption{The statistics of our RecFound dataset.}
\resizebox{\textwidth}{!}{%
\begin{tabular}{l ccc ccccccccccc c}
\toprule
\multicolumn{1}{c}{\multirow{3}{*}{Split}} & \multicolumn{3}{c}{Embedding Tasks} & \multicolumn{10}{c}{Generative Tasks} & \multirow{3}{*}{Total} \\
\cmidrule(lr){2-4} \cmidrule(lr){5-14}
& \multicolumn{3}{c}{/}
& \multicolumn{3}{c}{General NLP} & \multicolumn{4}{c}{User Understanding} & \multicolumn{3}{c}{Item Understanding} & \\
\cmidrule(lr){2-4}\cmidrule(lr){5-7} \cmidrule(lr){8-11} \cmidrule(lr){12-14}
& \multicolumn{1}{c}{U2I} & \multicolumn{1}{c}{Q2I} & \multicolumn{1}{c}{I2I} & QR & AVE & AG & SR & SA & UP & AP & PRP & PM & IP & \\
\midrule
Training   & 86,000 & 40,000 & 83,680 & 3,260 & 20,000 & 20,000 & 20,000 & 20,000 & 20,000 & 15,000 & 20,000 & 4,000 & 15,000 & 366,940 \\
Validation & 3,500  & 2,600  & 3,340  & 500   & 500    & 500    & 500    & 500    & 500    & 500    & 500    & 500   & 500    & 14,940 \\
Testing    & 4,800  & 4,000  & 4,800  & 124   & 1,334  & 1,332  & 1,334  & 1,334  & 227    & 1,334  & 1,333  & 168   & 266    & 22,383 \\
\bottomrule
\end{tabular}%
}
\label{tab:data}
\end{table}

\section{Methodology}

\subsection{General Training Pipeline}

Before introducing our specific methods, we first outline the general training pipeline used for both embedding and generative tasks. 
In our unified framework, all raw input data is transformed into standardized instruction-response pairs. 
We give example data for each task in Appendix~\ref{app:data}.
Then, these samples are processed via two branches for embedding and generative tasks, respectively. 

In the \textbf{embedding branch}, batched inputs are passed through a shared LLM backbone and trained using a \textit{contrastive objective} (with InfoNCE loss) to learn a well-structured embedding space.
When training backbone is a decoder-only model, we extract the embedding representation for each input sequence by applying mean pooling over the final layer hidden states.
In the \textbf{generative branch}, samples are similarly batched and optimized using a \textit{token-level cross-entropy loss} to minimize sequence generation errors. 
Since we aim to establish a unified recommendation foundation model, both branches share the same LLM backbone and parameters, with their respective gradients jointly backpropagated for model update.

However, the shared LLM backbone for multiple heterogeneous embedding and generative tasks inevitably introduces challenges of knowledge sharing \& conflict and convergence inconsistency. To address these issues, our RecFound framework incorporates three key modules: Task-wise Mixture of Low-Rank Experts, Step-wise Convergence-Oriented Sample Scheduler, and Model Merge.

\subsection{Task-wise Mixture of Low-Rank Experts}
\label{sec:task_tmole}

We propose Task-wise Mixture of Low-Rank Experts (\textbf{TMoLE}), which integrates multiple low-rank adapters~\cite{hu2021lora} with a task-wise routing mechanism. 
TMoLE layers are inserted in the multi-head attention projections-specifically, the query, key, value, and output transformations-across each Transformer layer.
Each base weight matrix \( W_0 \) of one projection is augmented with \( N \) parallel low-rank adapters and a \textbf{task-wise} router. 
Different from traditional token-level MoE approaches~\cite{luo2024moelora,dou2024loramoealleviateworldknowledge}, our router leverages the task type embedding associated with each training sample.
Given a task ID \( i \), it is mapped through a learnable embedding table to obtain the corresponding task embedding \( \mathbf{e}_i \).

The \( N \) low-rank experts are partitioned into three distinct groups: embedding experts \( \mathcal{E} \), generative experts \( \mathcal{G} \), and shared experts \( \mathcal{S} \). Each adapter \( j \) comprises trainable matrices \( A_j \in \mathbb{R}^{d \times r} \) and \( B_j \in \mathbb{R}^{r \times d} \), where \( r \ll d \). The task-wise router is implemented as a multi-layer perceptron (MLP) coupled with a masking module. The output gating vector \( v \) (i.e., the routing distribution) determines the contribution of each low-rank expert to the final TMoLE output, computed as:
\begin{equation}
    v_i = \mathrm{Softmax}(\mathrm{Mask}(\mathrm{MLP}(e_i))).
    \label{eq:routing distribution}
\end{equation}
Here, $\mathrm{Mask}(\cdot)$ suppresses irrelevant task dimensions by assigning them a value of \( -\infty \), thereby retaining only the experts pertinent to the current task and the shared experts. 
For example, the corresponding elements of embedding experts in $v_i$ are zero when routing for generative tasks.
Notably, the MLP is shared across all the tasks.

The final output of the TMoLE-enhanced projection layer for sample from task $i$ is given by:
\begin{equation}
    o = W_0 x + \sum\nolimits_{j=1}^{N} v_i  A_j B_j x .
\end{equation}
This architecture allows each expert to specialize based on task categories, while the shared experts facilitate cross-embedding-generative task generalization. 
The router and experts are trained jointly, promoting end-to-end task-specific adaptation. This design extends previous PEFT-MoE frameworks (e.g., LoRA-MoE~\cite{hu2021lora}, TaskMoE~\cite{kudugunta2021taskmoe}, and DoRA~\cite{liu2024dora}) and uniquely incorporates task-wise masking and gating over low-rank experts to resolve the problem of knowledge sharing \& conflict effectively.

\subsection{Step-wise Convergence-Oriented Sample Scheduler}

During the training, each task would exhibit different convergence speeds and different levels of difficulty, due to their task-specific characteristics, as well as the inherent knowledge possessed by LLM backbones. 
Existing works~\cite{gong2024coba,lin2017focal} propose adaptive loss reweighting according to the learning dynamics. 
However, we observe that direct loss reweighting leads to training instability, particularly when tasks use fundamentally different loss functions (i.e., InfoNCE for embedding tasks and Cross Entropy for generative tasks).

Hence, we propose Step-wise Convergence-Oriented Sample Scheduler (\textbf{S2Sched}), which estimates the non-convergence rate to schedule the proportion of samples for each task at each training step, rather than simply modifying loss weights. 
Specifically, we will first perform step-wise non-convergence estimation for each task, and then conduct task-level sample scheduling according to the estimation results, i.e., decreasing the sample ratio of a task if it is converging fast and vice versa.

\begin{figure}[t]
    \centering
    \includegraphics[width=\textwidth]{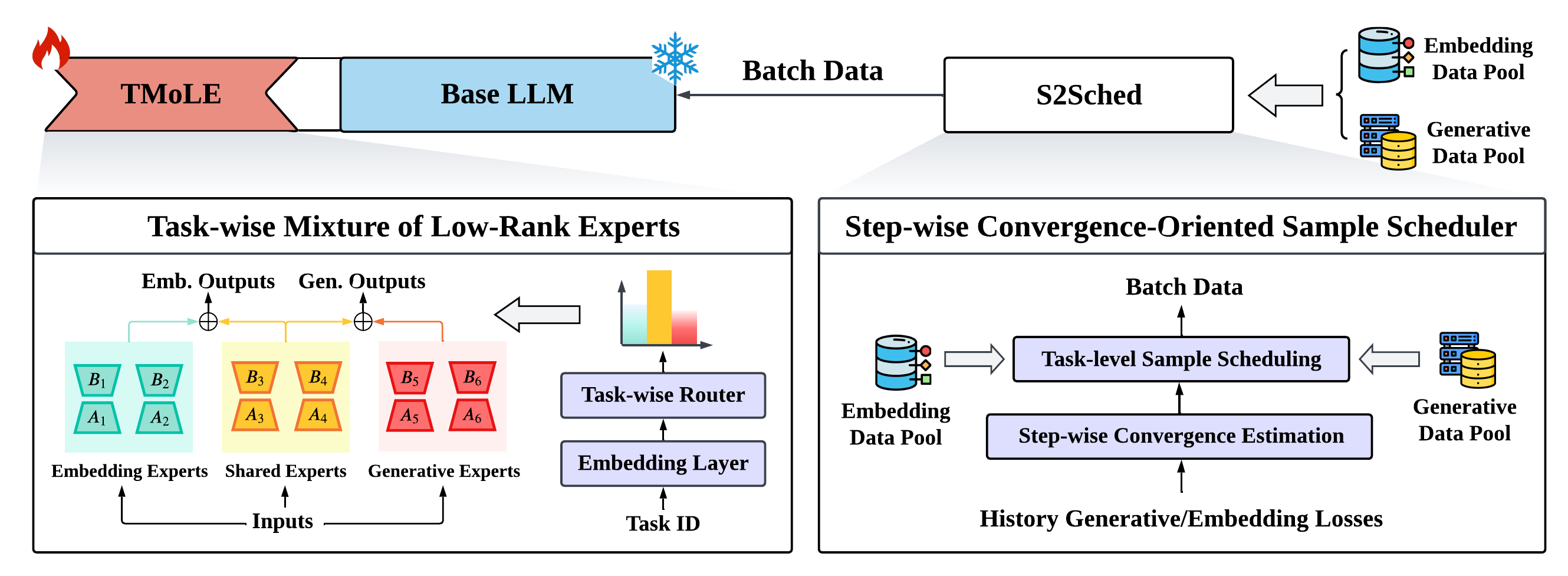}
    \caption{The Method of RecFound. In TMoLE structure, we assume we use 2 embedding experts, 2 shared experts, and 2 generative experts, respectively.}
    \label{fig:method}
\end{figure}

\subsubsection{Step-wise Non-Convergence Estimation}

We estimate the non-convergence rate based on the intuition that the validation loss slope of a nearly converged task should approach zero. 
It is worth noting that we decouple the estimation of embedding and generative tasks to avoid interference caused by differences in loss magnitudes. 
Hence, we denote the sets of embedding and generative tasks as $\mathcal{T}^{(\mathrm{E})}$ and $\mathcal{T}^{(\mathrm{G})}$, respectively. 

For each task type $b\in\{\mathrm{E},\mathrm{G}\}$ and task $i\in\mathcal{T}^{(b)}$ at training step $t$, we can calculate the normalized validation loss slope via linear regression over the last $L$ steps, denoted as $\alpha_i(t)$. 
Hence, we set a warmup ratio of $\varsigma$ to collect the initial validation loss sequences, i.e., we do not apply non-convergence estimation and sample scheduling at the warmup stage.
Based on the step-wise validation loss slopes, we estimate the non-convergence status through the following metrics:

\paragraph{Inter‑Task Non‑Convergence Rate.} 
We first consider the inter‑task  convergence status by comparing the tasks of the same type $b$, i.e., embedding or generative tasks:
\begin{equation}
  \gamma_{i,b}^{\mathrm{inter}}(t)
  = -\mathrm{Softmax}_i\!\left(
    |\mathcal{T}^{(b)}| \cdot \frac{\alpha_i(t)}{\sum_{i=1}^{|\mathcal{T}^{(b)}|} |\alpha_i(t)|}
  \right),
\end{equation}
where $\alpha_i(t)$ is the normalized validation loss slope of task $i$ estimated over the last $L$ steps. 
Higher $\gamma^{\mathrm{inter}}_{i,b}(t)$ indicates less convergence of task $i$ at step $t$ in comparison to other tasks.

\paragraph{Intra‑Task Non‑Convergence Rate.}
In addition to the inter-task non-convergence among tasks, we also estimate the intra-task non-convergence rate based on the recent $L$ training steps:
\begin{equation}
  \gamma_{i,b}^{\mathrm{intra}}(t)
  = \mathrm{Softmax}_i\!\left(
    -\,L \cdot \frac{\alpha_i(t)}{\sum_{s=t-L+1}^{t} |\alpha_i(s)|}
  \right).
\end{equation}
$\gamma_{i,b}^{\mathrm{intra}}(t)$ measures the stability of task $i$’s convergence by comparing the current slope to its own recent history. 
Higher $\gamma^{\mathrm{intra}}_{i,b}(t)$ indicates less convergence of task $i$ at step $t$ comparison to its history.

\paragraph{Balancing Weight.} To balance the inter- and intra-task non-convergence rate, we calculate the balancing weight as follows:
\begin{equation}
  \beta_b(t)
  = \min\;\Bigl\{
    t \cdot \mathrm{Softmax}_t\!\Bigl(
      -\,\frac{\tau t \max_i \alpha_i(t)}{\sum_{k=1}^{t}  \max_i |\alpha_i(k)|}
    \Bigr),\;1
  \Bigr\},
\end{equation}
where $\tau$ is temperature hyperparameter, and $\beta_b(t)$ is shared for all the tasks of type $b$. Higher $\beta_b(t)$ indicates a balanced relative converging speed among tasks at step $t$. Lower $\beta_b(t)$ indicates a fairly large disparity in convergence rates across tasks and potential abnormal values estimated by $\gamma_{i,b}^{\mathrm{inter}}(t)$.

\paragraph{Sample Ratio.} We interpolate the inter- and intra-task with the balancing weight to calculate the final sample ratio for each task $i$:
\begin{equation}
  \omega_{i,b}(t)
  = \mathrm{Softmax}_i\;\Bigl(
    - \beta_b(t)\,\gamma_{i,b}^{\mathrm{inter}}(t)
    + \bigl[1 - \beta_b(t)\bigr]\,\gamma_{i,b}^{\mathrm{intra}}(t)
  \Bigr),
  \label{eq:omega}
\end{equation}
where the terms $-\beta_b(t)\,\gamma_{i,b}^{\mathrm{inter}}(t)$ and $\bigl[1 - \beta_b(t)\bigr]\,\gamma_{i,b}^{\mathrm{intra}}(t)$ both convey the positive correlations with the overall non-convergent level for task $i$ of type $b$ at training step $t$.

\subsubsection{Task-level Sample Scheduling}

We maintain different batch sizes and schedulers for embedding and generative tasks. 
For each batch of size \(B^{(b)}\) of task type $b\in\{E,G\}$ at step $t$, the scheduler would re-allocate the sample budget for each task $i\in\mathcal{T}^{(b)}$ according to the noramlized task weight $\omega_{i,b}(t)$. 
Specifically, the number of samples of task $i$ at step $t$ is $n_i^{(b)}(t) = \left\lfloor B^{(b)} \cdot w_{i,b}(t) \right\rfloor$. 
Then, we would conduct uniform sampling without replacement from the task-wise training data pool to form the final training batch. 

In this way, S2Sched adaptively adjusts the task sample ratio per batch within different task types (embedding or generative) at each training step according to real-time validation feedback. 
Therefore, RecFound can achieve stable and convergence-balanced multi-task learning for recommendation foundation models.

\subsection{Model Merge}

In multi-task learning scenarios, different model checkpoints often specialize in particular tasks or objectives. This specialization interacts with the inherent seesaw effect of foundation model training, where performance improvements on one task may coincide with regressions on others. 
To mitigate such interference and consolidate complementary knowledge, we adopt TIES-Merging~\cite{yadav2023ties} as the Model Merge module over the learned low-rank experts across different checkpoints.
We first remove insignificant parameter updates, then resolve conflicting update directions via majority voting, and finally average the consistent parameters. 
This approach produces a merged model that better balances the performance across different tasks.

\begin{table}[t]
\centering
\caption{Performance on generative tasks. Average Rank is obtained by calculating the average of the model's rankings on each task. The smaller this value is, the better the comprehensive effect of the model will be. For each task, the best result is in \textbf{bold}, and the second-best result is \underline{underlined}.}
\resizebox{\textwidth}{!}{%
\begin{tabular}{lcccccccccc c}
\toprule
\multicolumn{1}{c}{\multirow{3}{*}{Models}} & \multicolumn{3}{c}{General NLP} & \multicolumn{4}{c}{User Understanding} & \multicolumn{3}{c}{Item Understanding} & \multicolumn{1}{c}{\multirow{2}{*}{\textbf{Average Rank}↓}} \\
\cmidrule(lr){2-4} \cmidrule(lr){5-8} \cmidrule(lr){9-11}
 & QR↑ & AVE↑ & AG↑ & SR↑ & SA↑ & UP↑ & AP↑ & PRP↑ & PM↑ & IP↑ & \\
 & (F1) & (F1) & (F1) & (R@1) & (F1) & (F1) & (F1) & (F1) & (F1) & (F1) & (Mean (Std)) \\
\midrule
\multicolumn{12}{c}{\textbf{Baseline Models w/o Finetuning}} \\
\cmidrule(lr){1-12}
Llama3-8B-Instruct & 0.926 & 0.000 & 0.830 & 0.000 & 0.061 & 0.875 & 0.420 & 0.231 & 0.444 & 0.880 & 9.30 (2.57) \\
Mistral-7B-Instruct & 0.936 & 0.000 & 0.841 & 0.005 & 0.408 & 0.873 & 0.738 & 0.103 & 0.449 & \underline{0.892} & 7.20 (2.71) \\
Mixtral-8x7B-Instruct & 0.948 & 0.000 & 0.840 & 0.005 & 0.278 & 0.871 & 0.647 & 0.316 & 0.047 & 0.885 & 8.20 (1.94) \\
Qwen2.5-7B-Instruct & 0.952 & 0.000 & 0.834 & 0.027 & 0.448 & 0.872 & 0.383 & 0.208 & 0.215 & 0.881 & 8.60 (1.80) \\
Qwen2.5-14B-Instruct & 0.944 & 0.000 & 0.832 & 0.063 & 0.468 & 0.872 & 0.518 & 0.251 & 0.434 & 0.886 & 7.50 (1.43) \\
Qwen2.5-32B-Instruct & 0.961 & 0.000 & 0.828 & 0.128 & 0.466 & 0.872 & 0.488 & 0.212 & 0.373 & 0.887 & 7.70 (2.24) \\
GritLM-7B & 0.960 & 0.000 & 0.840 & 0.000 & 0.227 & 0.868 & 0.285 & 0.219 & 0.491 & 0.881 & 8.70 (2.37) \\
eCeLLM-L & 0.943 & 0.002 & \underline{0.843} & 0.177 & 0.362 & 0.848 & 0.812 & 0.361 & \textbf{0.994} & 0.857 & 6.70 (3.61) \\
\cmidrule(lr){1-12}
\multicolumn{12}{c}{\textbf{Baseline Models w/ Finetuning}} \\
\cmidrule(lr){1-12}
eCeLLM & \underline{0.973} & 0.315 & 0.841 & \textbf{0.251} & 0.532 & \underline{0.878} & 0.821 & 0.450 & 0.967 & 0.888 & 3.20 (1.08) \\
GritLM & \underline{0.973} & 0.316 & 0.839 & \underline{0.242} & 0.501 & 0.874 & 0.814 & 0.426 & 0.967 & 0.889 & 4.10 (1.51) \\
\cmidrule(lr){1-12}
\multicolumn{12}{c}{\textbf{Our Models}} \\
\cmidrule(lr){1-12}
RecFound w/o Model Merge & \textbf{0.974} & \underline{0.341} & \textbf{0.853} & 0.234 & \textbf{0.545} & \textbf{0.895} & \underline{0.827} & \textbf{0.471} & \underline{0.988} & \underline{0.902} & \underline{1.80 (0.98)} \\
RecFound & \underline{0.973} & \textbf{0.344} & \textbf{0.853} & \textbf{0.251} & \underline{0.533} & \textbf{0.895} & \textbf{0.830} & \underline{0.462} & \textbf{0.994} & \textbf{0.903} & \textbf{1.30 (0.46)} \\
\bottomrule
\end{tabular}%
}
\label{tab:generative}
\end{table}

\begin{table}[t]
 \centering
 \caption{Performance on embedding tasks. Notations are the same as Table~\ref{tab:generative}.}
 \resizebox{0.75\textwidth}{!}{%
 \begin{tabular}{lccc c} 
 \toprule
 \multicolumn{1}{c}{\multirow{2}{*}{Models}} & Item Emb↑ & Query Emb↑ & User Emb↑ & \textbf{Average Rank}↓ \\ 
  & (NDCG@20) & (MRR@20) & (MRR@20) & (Mean (Std)) \\ 
 \midrule
 \multicolumn{5}{c}{\textbf{Baseline Models w/o. Finetuning}} \\ 
 \cmidrule(lr){1-5} 
 gte-Qwen2-7B-Instruct & 0.806 & 0.975 & 0.431 & 6.00 (0.00) \\
 SFR-Embedding-Mistral & 0.805 & 0.974 & 0.419 & 7.00 (0.00) \\
 bge-large-en-v1.5 & 0.799 & 0.957 & 0.419 & 8.33 (1.15) \\
 Llama-3-8B-Instruct & 0.703 & 0.438 & 0.251 & 10.00 (0.00) \\
 Mistral-7B-Instruct & 0.805 & 0.974 & 0.419 & 7.00 (0.00) \\
 GritLM-7B & \underline{0.823} & 0.985 & 0.458 & 2.67 (0.58) \\
 eCeLLM-L & 0.697 & 0.197 & 0.221 & 11.00 (0.00) \\
 \cmidrule(lr){1-5} 
 \multicolumn{5}{c}{\textbf{Baseline Models w/ Finetuning}} \\ 
 \cmidrule(lr){1-5} 
 Embed.Only & 0.819 & 0.983 & 0.447 & 4.67 (0.58) \\
 GritLM & 0.820 & 0.985 & 0.439 & 4.00 (1.00) \\
 \cmidrule(lr){1-5} 
 \multicolumn{5}{c}{\textbf{Our Models}} \\ 
 \cmidrule(lr){1-5} 
 RecFound w/o Model Merge & 0.821 & \underline{0.987} & \textbf{0.573} & \underline{2.00 (1.00)} \\
 RecFound & \textbf{0.824} & \textbf{0.988} & \underline{0.570} & \textbf{1.33 (0.58)} \\
 \bottomrule
 \end{tabular}%
 }
 \vspace{-7pt}
 \label{tab:embedding}
 \end{table}

\section{Experiments}

\subsection{Experimental Setup}

\paragraph{Datasets.}  
All experiments are conducted on RecFound dataset. 
The data statistic is reported in Table~\ref{tab:data}. 
The prompt templates for generative and embedding tasks are designed with a specific format. We illustrate the prompt template for each task in Appendix~\ref{app:data}.

\paragraph{Baselines.}

We first adopt a wide range of non-tuned foundation models as baselines. 
For generative tasks, we choose Llama-3-8B-Instruct~\cite{grattafiori2024llama3herdmodels}, Mistral-7B-Instruct~\cite{jiang2023mistral7b}, Mixtral-8x7b-Instruct~\cite{jiang2024mixtralexperts}, and Qwen-2.5-Instruct series~\cite{yang2024qwen2}(7B,14B,32B).
For embedding tasks, we select gte-Qwen2-7B-Instruct~\cite{li2023gte}, SFR-Embedding-Mistral~\cite{SFRAIResearch2024}, bge-large-en-v1.5~\cite{bge_embedding}, Llama-3-8B-Instruct~\cite{grattafiori2024llama3herdmodels}, and Mistral-7B-Instruct~\cite{jiang2023mistral7b}.
Moreover, we also investigate different finetuning strategies based on our RecFound dataset. 
We adopt GritLM~\cite{muennighoff2025grit} as the common finetuned baseline for both generative and embedding tasks, since it is a unified generative representational instruction tuning method. 
We also choose eCeLLM~\cite{peng2024eCeLLM} (tuned on generative tasks) as the finetuned baselines for generative tasks. 
Since eCeLLM does not support embedding task training, we develop another baseline that discards the generative tasks and is only finetuned on the embedding tasks, i.e., Embed.Only.

\paragraph{Evaluation Metrics.}  
During evaluation, we allow up to 2048 newly generated tokens. 
In generative tasks, we adopt F1 score as the metric for query rewriting, attribute value extraction, answer generation, user profile, sentiment analysis, answerability prediction, item profile, product relation prediction, and product matching, and use Recall@1 for sequential recommendation. 
In embedding tasks, we measure the item feature embedding task with NDCG@k, and evaluate user query embedding and user sequence embedding tasks with MRR@k. Following previous works~\cite{tamm2021quality,hou2022towards}, we set $k=20$.

\paragraph{Implementation Details.} 

Our backbone LLM is Mistral-7B-Instruct~\cite{jiang2023mistral7b}.
We finetune the backbone LLM on the RecFound dataset, using a total batch size of 2048 for embedding tasks and 1024 for generative tasks. 
We use AdamW~\cite{loshchilov2017decoupled} with a learning rate of $2\!\times\!10^{-5}$. 
For TMoLE, we apply LoRA~\cite{hu2021lora} to the query/key/value/output projections with rank $r=16$, $\alpha=64$, and dropout rate 0.1. 
We instantiate $N=6$ low-rank experts for each projection, with the number of embedding experts $E=2$, generative experts $G=2$, and shared experts $S=2$. 
The task embedding size is 512. 
For S2Sched, we set the warmup ratio $\varsigma$ as 10\%, the length of the history window $L$ as 64 (last steps), and temperature $\tau$ as 10. 
We uniformly sample 128 validation data instances to calculate the normalized validation loss for each task at each training step.

\subsection{Main Results}

We evaluate our proposed RecFound and other baseline models. The results of generative and embedding tasks are reported in Table~\ref{tab:generative} and Table~\ref{tab:embedding}, respectively. We can observe that RecFound achieves the best performance across most generative tasks and embedding tasks, outperforming both larger-scale LLMs and strong instruction-tuned baselines. This demonstrates the effectiveness of TMoLE and S2Shed, which helps resolve the knowledge sharing \& conflict and convergence inconsistency problems for multi-task foundation model training.
In addition, the model merge module further improves the performance of RecFound for both generative and embedding tasks on average. 
This indicates that the model merge module combines the advantages of multiple model checkpoints, compensating for their limitations and showing excellent compatibility with heterogeneous multi-task learning.

\subsection{Ablation Study}

\begin{table}[t]
\centering
\caption{Ablation experiments on different model variants. M.M. stands for the Model Merge module.}
\resizebox{0.6\textwidth}{!}{%
\begin{tabular}{lcc}
\toprule
Model Variant & Generative (Avg.) & Embedding (Avg.) \\
\midrule
RecFound  & \textbf{0.704} & \textbf{0.775} \\
w/o M.M & 0.703 & \textbf{0.775} \\
w/o M.M. \& S2Sched   & 0.679 & 0.762 \\
w/o M.M. \& TMoLE  & 0.678 & 0.759 \\
w/o M.M. \& S2Sched \& TMoLE   & 0.684 & 0.748 \\
\bottomrule
\end{tabular}%
}
\vspace{-7pt}
\label{tab:ablation}
\end{table}

We conduct ablation studies by removing the Model Merge, S2Shed, and TMoLE components. The average results on generative and embedding tasks are reported in Table~\ref{tab:ablation}, while detailed results on each task are given in Appendix~\ref{app:ablation}. 
From Table~\ref{tab:ablation}, we can obtain the following observations:
\begin{itemize}
    \item Removing the model merge module only results in slight performance degradation. 
    While the model merge module leads to performance gains on some tasks and declines on others, it contributes to an overall improvement in model performance as shown in Table~\ref{tab:generative} and Table~\ref{tab:embedding}.
    \item When removing S2Shed, RecFound reverts to uniform sample scheduling among tasks. This leads to an imbalance of overfitting on simple tasks and underfitting on more challenging ones, thereby leading to inferior performance.
    \item Removing TMoLE reduces the learning structure of RecFound into a common LoRA~\cite{hu2021lora} module. This impairs the knowledge sharing and conflict resolution, which in turn degrades both generative and embedding performance significantly. 
\end{itemize}
These results confirm that Model Merge, S2Shed, and TMoLE are all essential for knowledge-disentangled and convergence-balanced multi-task learning of recommendation foundation models.

\subsection{In-Depth Analysis}

\subsubsection{Routing Distributions Among Tasks}

\begin{figure}[t]
    \centering
    \includegraphics[width=0.97\textwidth]{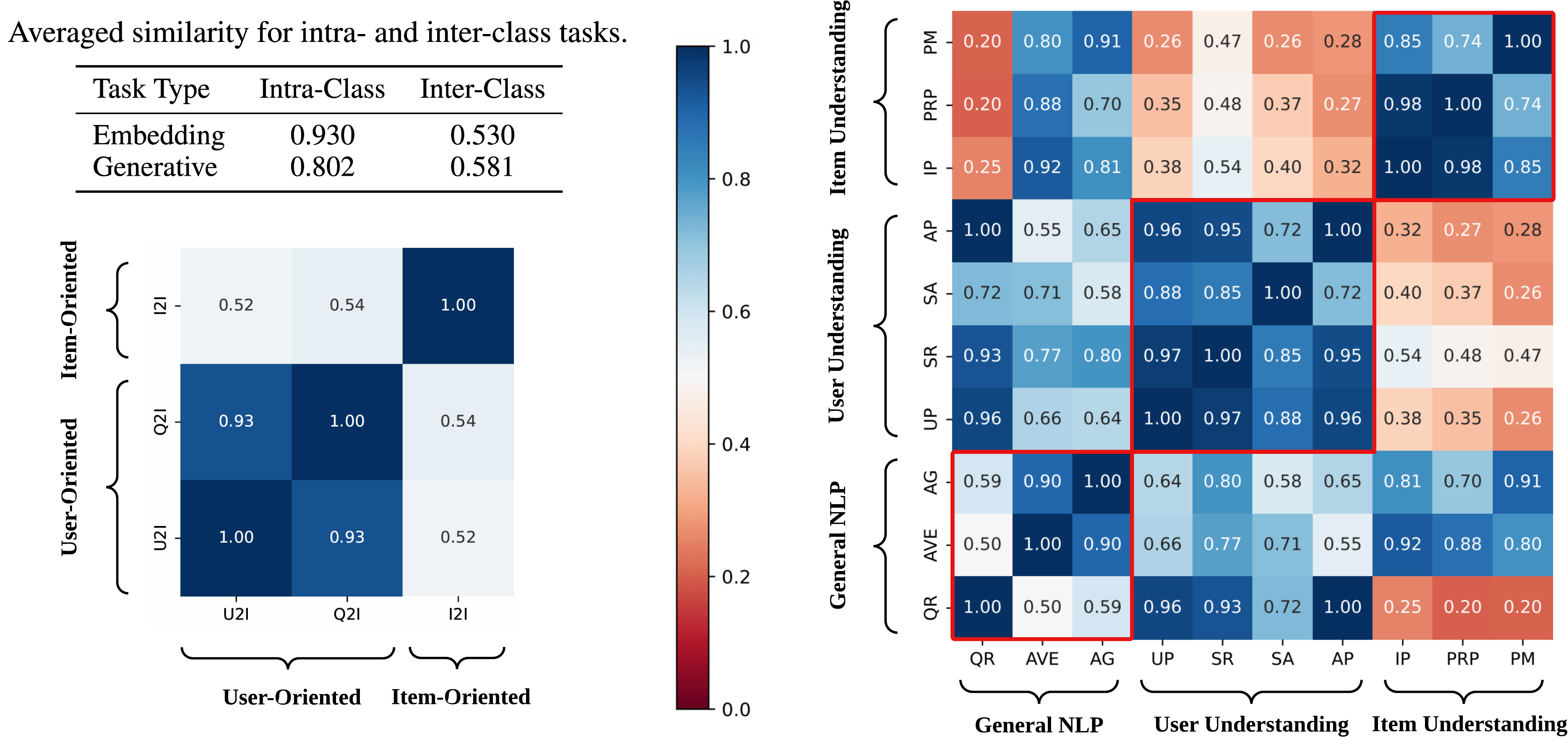}   
    \caption{
    Similarity heatmaps of the routing distribution across different tasks. 
    }
    \label{fig:sim}
\end{figure}

To investigate whether the router in TMoLE learns domain-specific characteristics for task-wise knowledge sharing and conflict, we visualize the similarity heatmap of the routing distribution $v_i$ in Eq.~\eqref{eq:routing distribution} across different tasks. 
Specifically, we compute the pairwise cosine similarity between each embedding or generative task pair, and illustrate the heatmap in Figure~\ref{fig:sim}.

We observe that tasks with higher similarity (i.e., intra-class denoted by the red box) tend to share more similar routing distributions, whereas tasks that differ substantially-particularly those from different classes-exhibit more divergent routing behaviors. 
This suggests that TMoLE can successfully capture both the commonalities and distinctions across tasks, thereby facilitating effective embedding \& generative task learning for recommendation foundation models.

\subsubsection{Scheduled Sample Ratio per Batch}

\begin{figure}[t]
    \centering
    \includegraphics[width=0.97\textwidth]{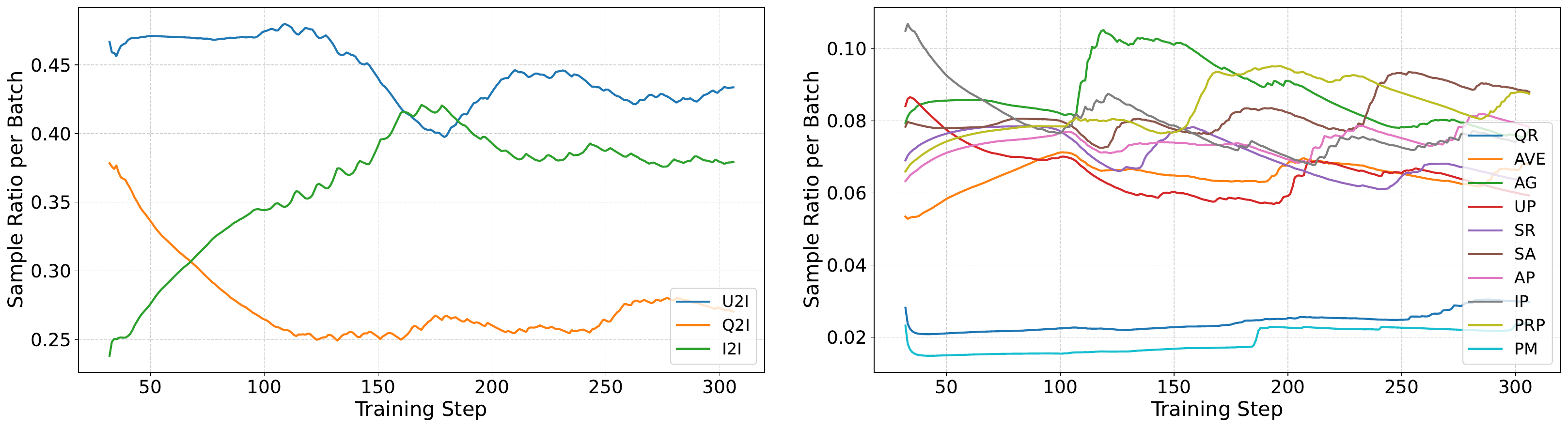}
    \caption{The evolving sample ratio per batch of different tasks during the training. The training step starts from 32 due to the warmup progress of S2Sched.}
    \vspace{-7pt}
    \label{fig:trend}
\end{figure}

To investigate how S2Sched works to balance the convergence speeds of different tasks, we visualize the sample ratio of different tasks per batch calculated in Eq.~\eqref{eq:omega} w.r.t. the training steps in Figure~\ref{fig:trend}. 
We observe that, during the training, the sample ratios of all tasks exhibit a general pattern of rising and falling alternately over time, rather than monotonically increasing/decreasing or remaining stable. 
A task that is approaching convergence will be assigned a lower weight, but later its weight will be increased again to prevent the forgetting problem of model learning. 
Such behavior reflects S2Sched’s ability to adaptively modulate sampling priorities across tasks, promoting more balanced optimization when tasks converge at different rates.

\section{Conclusion and Limitation}

In this work, we present RecFound, a unified generative representational learning framework for recommendation foundation models. 
We construct the first comprehensive dataset for recommendation foundation models covering 3 embedding tasks and 10 generative tasks. 
RecFound addresses the problems of knowledge sharing \& conflict and convergence imbalance through the proposed TMoLE, S2Sched, and Model Merge modules. 
Extensive experiments demonstrate strong generalization and task adaptability of RecFound model, highlighting the potential of unified recommendation foundation models for both generative and embedding tasks.
While effective, RecFound's performance may be sensitive to task-specific data imbalance and does not yet support continual task expansion. 
Future work will explore dynamic task adaptation, scalable routing, and extensions to multilingual and cross-modal recommendation scenarios.


\bibliographystyle{plain}
\bibliography{references}

\newpage
\appendix

\section{Broader Impact}
As a unified generative representational learning framework, RecFound promises to enhance personalization by capturing richer user–item interactions and broader tasks from various recommendation domains. 
It can improve recommendation quality, user satisfaction, and economic inclusivity for small businesses and content providers. 
However, recommendation foundation models can inadvertently amplify biases-leading to unfair exposure of items or user groups-and raise serious privacy concerns, as sensitive attributes may be inferred from learned embeddings.
To mitigate these risks, we recommend adopting privacy‑preserving techniques such as differential privacy or federated learning during data collection and model finetuning. 
Embedding human‑in‑the‑loop monitoring and user feedback mechanisms can further detect and correct unintended biases or misleading generative outputs, steering RecFound toward ethical and equitable recommendation practices.

\section{Data Construction Details}
\label{app:data}

\subsection{Prompt Templates}

The following prompts are used for evaluating general-purpose recommendation LLMs on generative tasks in RecFound Datasets.
For Mistral-7B Instruct-v0.3, the prompt is wrapped with "[INST]".
For Llama-3-8B-Instruct and Qwen-2.5-Instruct series, the prompt is wrapped with "<im\_start>" and "<im\_end>". 
We conduct zero-shot evaluations by default, so there's no one-shot or few-shot prompt template.

\paragraph{Prompt Template with Options for Generative Tasks}
\begin{itemize}
    \item System prompt: Below is an instruction that describes a task. Write a response that appropriately completes the request.
    \item Instruction: \{instruction\}
        \begin{itemize}
            \item input: \{input\}
            \item options: \{options\}
            \item response: 
        \end{itemize}
\end{itemize}
 
\paragraph{Prompt Template without Options for Generative Tasks}
\begin{itemize}
    \item System prompt: Below is an instruction that describes a task. Write a response that appropriately completes the request.
    \item Instruction: \{instruction\}
        \begin{itemize}
            \item input: \{input\}
            \item response: 
        \end{itemize}
\end{itemize}

To train and evaluate embedding tasks, we need to add special tokens "<|embed|>" as bos token for embedding inputs. An embed EOS is useless as there is no generative loss on it, so it won't be learned. There is also no need to design a prompt template for embedding tasks, we only need to extract embedding vectors from the mean pooling of last-hidden states.

\subsection{Diverse  Instructions}
For each embedding and generative task, we design several instructions to help LLMs understand various user requests. For each task, we show all the instructions we designed below.


\paragraph{Query Rewriting}
\begin{itemize}
\item Please undertake the task of rephrasing the questions and responses in a given conversation. You will be given a dialogue between two individuals (I will use A: for the first person and B: for the second person). For each query/response, an indication will be provided in parentheses at the end showing whether it requires rewriting. If no rewriting is needed, simply return the original content; if it does, use the preceding context to modify the sentence appropriately, making its meaning clearer, such as by specifying pronouns, adding place names or times, etc. 
\item Please carry out a task to rephrase the queries and replies in a given conversation. You will receive a dialogue between two people (I will label the first person as A: and the second person as B:). For each query/reply, I will note in parentheses at the end whether the sentence needs rewriting. If no rewriting is required, simply return the original text; if rewriting is required, use the preceding context to modify the sentence appropriately to make its meaning clearer, such as specifying pronouns, adding locations or times, etc. 
\item Please accomplish the task of rephrasing the queries and responses in a given conversation. You will be provided with a dialogue between two individuals (I will use A: to refer to the first person and B: to refer to the second person). For each query/response, I will specify in parentheses at the end whether it needs to be rewritten. If no rewriting is needed, return the original content; if rewriting is required, use the previous context to appropriately modify the sentence to make its meaning clearer, such as specifying pronouns, adding place names, or times, etc.
\item Please perform the task of rephrasing the questions and answers in a given conversation. You will receive a dialogue between two people (denoted as A: for the first person and B: for the second person). For each question/answer, I will indicate at the end in parentheses whether it needs to be rewritten. If no rewriting is necessary, simply return the original content; if rewriting is necessary, use the context to modify the sentence appropriately to make its meaning clearer, such as by specifying pronouns, adding place names, or times, etc. 
\item Please complete a task of rewriting the queries and replies in a given conversation. You will receive a dialogue between two people (I will use A: to denote the first person and B: to denote the second person). For each query/reply, I will indicate in parentheses at the end whether the sentence needs to be rewritten. If rewriting is not required, you just need to return the original content; if rewriting is required, please use the preceding context to modify the sentence appropriately to make its meaning clearer, such as specifying pronouns, adding place names or times, etc.
\end{itemize}

\paragraph{Attribute Value Extraction}
\begin{itemize}
\item Given the title, description, feature, price, and brand of a product and a set of target attributes, extract the value of each target attribute from the product information. Output the extracted value and the corresponding source (e.g., title or feature) denoting where the value is extracted.
\item Extract the value of the target attribute from the given product information and output it along with the corresponding source.
\item Parse the product information to locate the target attribute, and then provide the extracted value of the target attribute and its source in the output, specifying None if the attribute is not present.
\item First, identify the target attributes from the provided list. Then, scan the product title, description, feature, and brand to extract the values associated with each target attribute. Finally, create a list of dictionaries, each containing the extracted attribute, its corresponding value, and the source where it was found.
\item Using the product's title, description, features, price, and brand, identify and retrieve the values associated with a specified set of target attributes. Output the extracted values along with their respective sources (e.g., title or feature) indicating where each value was found. 
\end{itemize}

\paragraph{Answer Generation}
\begin{itemize}
\item Given a question and the related document, and generate the answer to the question based on the information provided in the document.
\item Generate an answer to the question by utilizing the information contained in the document.
\item Extract information from the supporting document to answer the given question.
\item Answer the given question using the supporting document.
\item Answer the given question by extracting information from the supporting document.
\end{itemize}

\paragraph{User Profile}
\begin{itemize}
\item Given a [Male/Female] user who is aged 18-24 and in a [student/service staff/.../unknown occupation], this user's movie viewing history over time is listed below. [History Sequence] Analyze user's preferences on movies (consider factors like genre, director/actors, time period/country, character, plot/theme, mood/tone, critical acclaim/award, production quality, and soundtrack). Provide clear explanations based on relevant details from the user's movie viewing history and other pertinent factors.
\item Given user's book rating history: [List of Bookname: Rating] Analyze user's preferences on books about factors like genre, author, writing style, theme, setting, length and complexity, time period, literary quality, critical acclaim (Provide clear explanations based on relevant details from the user's book viewing history and other pertinent factors.)
\end{itemize}

\paragraph{Sequential Recommendation}
\begin{itemize}
\item Given the products the user has purchased in history, rank the items in the listed options and output the item that the user is most likely to purchase next. Answer from one of the options.
\item Based on the user’s historical purchases, rank the items in options and predict the next product of the user’s interest from the provided options.
\item Rank the items in options and predict the user's next purchase from the listed options by analyzing her historical purchases.
\item The user's purchase history implies her preferences. Rank the items in the options based on the user’s preferences. Output the item that the user is most likely to purchase next from the options.
\item Rank items in listed options based on the user’s purchase history to determine the item that the user is most likely to purchase next. Output the item with the highest likelihood of being the next purchase.
\end{itemize}

\paragraph{Sentiment Analysis}
\begin{itemize}
\item Given the user's review, identify the user's sentiment from the listed options. Answer using one of the options.
\item Assess the user's sentiment in the provided review and select the appropriate sentiment option from the list as the answer.
\item Determine the sentiment expressed by the user in her review from the provided choices, and respond by selecting one of the available options.
\item Carefully assess the user's review for any strong expressions of sentiment, either positive or negative. Based on your analysis, select the most fitting sentiment option from the provided list as output.
\item Analyze the user's review text and determine the overall sentiment expressed, then choose the corresponding sentiment option from the provided list (A: very positive, B: positive, C: neutral, D: negative, E: very negative) based on the identified sentiment.
\end{itemize}

\paragraph{Answerability Prediction}
\begin{itemize}
\item Given a question and the related document, predict if the question is answerable based on the information provided in the document. Output only yes or no.
\item Evaluate the answerability of a question by analyzing the related document, outputting yes if the document contains information addressing the question, and no otherwise.
\item Analyze a question and its supporting document. Predicting answerability based on the information provided in the document. Output yes if the document contains relevant information to answer the question, otherwise output no.
\item Given a question and its related document, determine if the question is answerable by analyzing the information in the document. Output yes if the document addresses the question, or no otherwise.
\item Output yes if the supporting document can answer the given question. Otherwise, output no.
\end{itemize}

\paragraph{Item Profile}
\begin{itemize}
\item Introduce movie [Movie name] and describe its attributes (including but not limited to genre, director/cast, country, character, plot/theme, mood/tone, critical acclaim/award, production quality, and soundtrack).
\item Introduce book [Book name], which is from [Source] Page and describe its attributes including but not limited to genre, author, writing style, theme, setting, length and complexity, time period, literary quality, critical acclaim.
\end{itemize}

\paragraph{Product Relation Prediction}
\begin{itemize}
\item Given the title of two products, predict if the two products are similar, if the two products will be purchased or viewed together. Answer only from the options.
\item Analyze the titles of Product 1 and Product 2 to determine if they are similar, if they will be purchased or viewed together, and choose the corresponding option.
\item Evaluate the titles of Product 1 and Product 2, then choose the option that best describes the relation between the two products.
\item Evaluate the titles of Product 1 and Product 2 to assess their similarity and whether they are likely to be purchased or viewed together. Then, select the appropriate option.
\item Predict whether two products are similar, whether two products are likely to be purchased or viewed together based on their titles. Choose your answer from the provided options.
\end{itemize}

\paragraph{Product Matching}
\begin{itemize}
\item Given the title, description, manufacturer, and price of two products, identify if they are the same product. Only output yes or no.
\item Analyze the title, description, manufacturer, and price between the two products below and generate an output of yes if the two products are the same, otherwise respond with no.
\item Check the details of the two products to see if they refer to the same product. Output only yes or no.
\item Based on the product information, predict if the two products are identical or not. Output yes if they are identical or no otherwise.
\item Compare the details of two given products to determine if they are identical. Output yes if they are identical or no otherwise.
\end{itemize}

\paragraph{User2Item}
\begin{itemize}
\item Represent this sequence of user actions to predict the next product he/she will purchase (User sequence is in reverse chronological order):
\item Encode this user activity sequence to forecast their next product purchase (User sequence is in reverse chronological order):
\item Create an embedding of this user sequence to determine the next product they will purchase (User sequence is in reverse chronological order):
\item Generate a representation of this user sequence to find the next product this user will buy (User sequence is in reverse chronological order):
\item Represent this user sequence to retrieve the next product this user will buy (User sequence is in reverse chronological order):
\end{itemize}

\paragraph{Query2Item}
\begin{itemize}
\item Represent this user query to find a matching item:
\item Generate a representation of this user query to locate a relevant item:
\item Create an embedding of this user query to retrieve a relevant item:
\item Use the representation of the user query to identify a relevant item:
\item Represent this user query to retrieve a relevant item:
\end{itemize}

\paragraph{Item2Item}
\begin{itemize}
\item Create an embedding for the product using its title and categories to help find similar items
\item Generate an embedding for a product based on its title and categories to facilitate the retrieval of similar items
item Based on the title and categories of a product, generate an embedding to identify similar products
\item Using the product's title and categories, produce an embedding to retrieve items that are alike
\item Given the title of a product and its category, please generate an embedding for the product to retrieve similar items
\end{itemize}

\subsection{Data Examples}
In this section, we provide an example for each task, with complete instructions (including system prompt and user prompt) and the corresponding ground-truth.

\paragraph{Query Rewriting}
\begin{itemize}
\item
\textbf{Instruction:} 

Please perform the task of rephrasing the questions and answers in a given conversation. You will receive a dialogue between two people (denoted as A: for the first person and B: for the second person). For each question/answer, I will indicate at the end in parentheses whether it needs to be rewritten. If no rewriting is necessary, simply return the original content; if rewriting is necessary, use the context to modify the sentence appropriately to make its meaning clearer, such as by specifying pronouns, adding place names, or times, etc. Now the dialogue you need to rewrite is shown below: 

A: Do I have missed calls today? (rewrite required: False)

B: Yes, you have two. It's from Don and Ezra. (rewrite required: True)

A: Call Ezra first. (rewrite required: False)

B: Calling her now. (rewrite required: True)

Based on the given dialogue, your rewritten result is:
\item
\textbf{Answer:}

A: Do I have missed calls today?

B: Yes, you have two missed calls today. It's from Don and Ezra.

A: Call Ezra first.

B: Calling Ezra now.
\end{itemize}

\paragraph{Attribute Value Extraction}
\begin{itemize}
\item 
\textbf{Instruction:} 

Below is an instruction that describes a task, paired with an input that provides further context. Write a response that appropriately completes the request.

--Instruction:

Given the title, description, feature, price, and brand of a product and a set of target attributes, extract the value of each target attribute from the product information. Output the extracted value and the corresponding source (e.g., title or feature) denoting where the value is extracted.

--Input:

Product title: Bencore Multi Functional Molle Tactical Messenger Bag.

Product description: This rugged/durable tactical shoulder bag provides perfect and stylish solution for almost any scenario. The bag is made of durable nylon construction that will not tear, color will not fade. The bag has many MOLLE straps through the bag for all your MOLLE accessories. The bag contains many roomy compartments as pictured and comes in many stylish colors. Design, comfort and functionality was the emphasis of this bag which is why we made sure the bag is fully ergonomic, lightweight and has many roomy pockets and Velcro patches throughout the bag. The product comes with the Bencore Life Time warranty and is satisfaction guaranteed. Bencore is a leading manufacturer in outdoor Apparel/Accessories, from Par cords to Backpacks to basic outdoor essentials.

Product feature: Durable heavy-duty, lightweight Nylon construction, will not tear or break even under extreme conditions - Lifetime Warranty, Rugged, roomy main drawstring-closed compartment provides secure storage space for your gear; MOLLE System, works with most MOLLE accessory, Front pocket provides quick access, roomy interior pocket for convenient separated storage, concealed back pocket with zipper closure, Padded and fully ergonomic System, adjustable shoulder strap for comfortable handling.

Product brand: Bencore

Target attributes: Material

--Response:

\item \textbf{Answer:}

Attribute: material; Value: nylon; Source: product description

Attribute: material; Value: nylon; Source: product feature.

\end{itemize}

\paragraph{Answer Generation}
\begin{itemize}
\item \textbf{Instruction:}

Below is an instruction that describes a task, paired with an input that provides further context. Write a response that appropriately completes the request.

--Instruction:

Generate an answer to the question by utilizing the information contained in the document.

--Input:

Would this be considered a dark metal pan which requires a lower baking temp than shiny pans?" 

Document: 

I've been baking all our bread for several years and needed new pans. These are very nice, a good weight, smooth finish, easy to clean, and the bread bakes perfectly. Would definitely recommend.

Great for baking bread. Bread slides out of Non-stick pan easily. Definitely would recommend this for baking. Sturdy design. I will be purchasing another one soon.", "I kept buying bread pans as my recipes called for an 8 by 4 bread pan. They would arrive and I would measure them and they were never four inches deep. Then I finally realized they were talking about width.

These are good metal pans just as the ad showed. There are not any flaws or other negatives. thanks, Mike", "I needed to purchase an 8x4 loaf pan for baking gluten free bread. It has been wonderful! My family is able to enjoy sandwiches again. The bread releases well and the pans wash easily. I do spray them with cooking spray before baking.

--Response:

\item \textbf{Answer:}

The pan is a dark metal, but I use the same temp as for lighter pans and bake until the top of my bread is golden brown and the crust within the pan seems to be just fine.
\end{itemize}

\paragraph{User Profile}

\begin{itemize}
\item \textbf{Instruction:}

Below is an instruction that describes a task. Write a response that appropriately completes the request.

--Instructions:

Given a female user who is aged 50-55 and self-employed, this user's movie viewing history over time is listed below. Sabrina (1995), 4 stars; Sabrina (1954), 5 stars; Gone with the Wind (1939), 5 stars; Citizen Kane (1941), 4 stars; You've Got Mail (1998), 3 stars; Thelma \& Louise (1991), 3 stars; American Beauty (1999), 4 stars; Broadway Melody, The (1929), 4 stars; American President, The (1995), 4 stars; Corrina, Corrina (1994), 4 stars; Analyze user's preferences on movies (consider factors like genre, director/actors, time period/country, character, plot/theme, mood/tone, critical acclaim/award, production quality, and soundtrack). Provide clear explanations based on relevant details from the user's movie viewing history and other pertinent factors.

--Response:

\item \textbf{Answer:}

Based on the user's movie viewing history, it seems that she enjoys classic films from the 1930s and 1940s, as well as more recent dramas and romantic comedies. She seems to appreciate films with strong female characters, as evidenced by her ratings of Thelma \& Louise and Sabrina (both versions). 

She also appears to value critical acclaim and award-winning films, as she gave high ratings to Gone with the Wind and Citizen Kane. Additionally, she enjoys films with good production quality, as indicated by her positive rating for The American President.

It is possible that the user has a preference for films with a romantic or sentimental tone, as her ratings for You've Got Mail, Corrina, Corrina, and the two versions of Sabrina are all relatively positive. 

Based on this information, it could be recommended that the user explore classic films from the 1930s and 1940s, as well as critically acclaimed dramas and romantic comedies. She may also enjoy films with strong female characters and a sentimental tone.

\end{itemize}

\paragraph{Sequential Recommendation}
\begin{itemize}
\item \textbf{Instruction:}

Below is an instruction that describes a task, paired with an input that provides further context. Write a response that appropriately completes the request.

--Instruction:

Given the products the user has purchased in history, rank the items in the listed options and output the item that the user is most likely to purchase next. Answer from one of the options.

1st: Wilson Pro feel Tennis Vibration Dampener. Sports \& Outdoors. Sports \& Fitness. Wilson.

2nd: COSMOS 5 PCS White cotton sports basketball headband / sweatband head sweat band/brace. Sports \& Outdoors. Sports \& Fitness. Cosmos.

3rd: Gamma Guard. Sports \& Outdoors. Sports \& Fitness. Gamma.

4th: SummerHouse 1 Gram Per Inch Golf High Density Lead Tape Weight Self-Adhesion for Driver Fairway Hybrid Iron Clubs Head - 1/2''x 100''. Sports \& Outdoors...

--Options:

A: Aceson A7 7Quad Core Google Android 4.4 Tablet PC, IPS Display, 8GB Storage, Bluetooth 4.0, Dual Camera, Support Netflix, Skype, Facebook, YouTube, Google...

B: Pinzon Heather Flannel Blanket, Full/Queen, Blue Heather. Home \& Kitchen. Bedding. Pinzon by Amazon

C: Sony LCS-SC5 Alpha System Carrying Case. Electronics. Camera \& Photo. Sony.

D: Advent Candles Set 4, 10" X 3/4". Home \& Kitchen. Home Dcor. Abbey Gift.

E: Nidec Servo Gentle Typhoon D1225C12B5AP 120mm 1850 rpm Silent Case Fan. Electronics. Computers \& Accessories. SERVO.

F: FYY Case for iPad Air 2 - Premium PU Leather Case Smart Auto Wake/Sleep Cover with Hand Strap, Card Slots, Pocket for iPad Air 2.

--Response:

\item \textbf{Answer:}

C

\end{itemize}

\paragraph{Sentiment Analysis}
\begin{itemize}
\item \textbf{Instruction:}

Below is an instruction that describes a task, paired with an input that provides further context. Write a response that appropriately completes the request.

--Instruction:

Given the user's review, identify the user's sentiment from the listed options. Answer using one of the options.

--Input:

This gives me control of the volume while on a treadmill and I don't have to stop and go back to the stereo or TV while I'm walking.

Options:

A: very positive

B: positive

C: neutral

D: negative

E: very negative

--Response:

\item \textbf{Answer:}

A: very positive

\end{itemize}

\paragraph{Answerability Prediction}
\begin{itemize}
\item \textbf{Instruction:}

Below is an instruction that describes a task, paired with an input that provides further context. Write a response that appropriately completes the request.

--Instruction:

Given a question and its related document, determine if the question is answerable by analyzing the information in the document. Output yes if the document addresses the question, or no otherwise.

--Input:

question: Is the lamp base weighted well?  Will the lamp tip or lean over if you adjust the light to any angle or extend the arm out to it's maximum capacity?, 

document: 
This is a wonderful light. It provides exactly the kind of light I need for the close hand work I do. I highly recommend this light!

This is my second lamp of this style and model. i purchased the first one locally for my husband several months ago. I paid almost \$60.00 more for that one. It was so helpful for him, reducing eye strain and thus allowing him to read more, that I decided to try to find another one for myself. I thought I was OK reading with just the regular lamps in our living room until I sat under my husband's Ott Lite. What a difference it made!

I understand it is an investment, but you only have one pair of eyes and the older you get, the harder it is to focus without eye strain and the consequential headaches that can accompany poor lighting. I'm glad I purchased these lamps and highly recommend them to anyone who loves to read.

--Response:

\item \textbf{Answer:}

No

\end{itemize}

\paragraph{Item Profile}
\begin{itemize}
\item \textbf{Instruction:}

Below is an instruction that describes a task. Write a response that appropriately completes the request.

--Instruction:

Introduce movie Raising Arizona (1987) and describe its attributes (including but not limited to genre, director/cast, country, character, plot/theme, mood/tone, critical acclaim/award, production quality, and soundtrack).

--Response:

\item \textbf{Answer:}

Raising Arizona is a comedy film directed by the Coen brothers and released in 1987. The main cast includes Nicolas Cage, Holly Hunter, and John Goodman. The film was produced in the United States.

The plot of Raising Arizona is centered around a couple, played by Cage and Hunter, who are unable to have children of their own. They decide to kidnap one of a wealthy family's quintuplets, but soon face trouble as the child's biological parents hire a bounty hunter to track them down.

The movie's tone is quirky and absurdist, with elements of slapstick comedy and witty dialogue. The soundtrack features a mix of country music and traditional film score.

Raising Arizona received critical acclaim upon its release and has since become a cult classic. The film's unique blend of humor, character development, and visual style has made it a beloved favorite among fans of the Coen brothers' work. The movie was nominated for several awards, including a Golden Globe for Best Performance by an Actor in a Motion Picture - Musical or Comedy for Nicolas Cage's performance.

\end{itemize}

\paragraph{Product Relation Prediction}
\begin{itemize}
\item \textbf{Instruction:}

Below is an instruction that describes a task, paired with an input that provides further context. Write a response that appropriately completes the request.

--Instruction:

Given the title of two products, predict if the two products are similar, if the two products will be purchased or viewed together. Answer only from the options.

--Input:

Product 1:: Gamo Whisper Fusion Pro 0.22-Calibre Air Rifle with 3-9x40 Adjustable Objective Scope, 

Product 2:: RWS Model 34 .22 Caliber Pellet Air Rifle Combo

--Options:

A: Users who buy product 1 may also buy product 2.

B: Users who view product 1 may also view product 2.

C: The product 1 is similar with the product 2.

--Response:

\item \textbf{Answer:}

"B"
\end{itemize}

\paragraph{Product Matching}
\begin{itemize}

\item \textbf{Instruction:}

Below is an instruction that describes a task, paired with an input that provides further context. Write a response that appropriately completes the request.

--Instruction:

Given the title, description, manufacturer, and price of two products, identify if they are the same product. Only output yes or no.

--Input:

"product 1": \{"title": "spycatcher 2007for pc", "description": NaN, "manufacturer": "avanquest", "price": "19.99"\}, 

"product 2": \{"title": "avanquest usa llc spycatcher 2007", "description": "detect and remove spyware spycatcher is the first and only anti-spyware solution that continuously protects consumers from next-generation mutating spyware easily and safely. spycatcher enables even novice pc users to remove and block the most ins", "manufacturer": "nan", "price": "17.5"\}

--Response:

\item \textbf{Answer:}

yes

\end{itemize}

\paragraph{User2Item}
\begin{itemize}

\item \textbf{Query:} 

Generate a representation of this user sequence to find the next product this user will buy (User sequence is in reverse chronological order):

Rawlings Players Series Youth T-Ball Glove, Regular, Basket-Web, 9 Inch

ProsourceFit Flex Foam Rollers

WeeRide Kangaroo Child Bike Seat

FOCO NHL Unisex Team Thematic Gnome

\item \textbf{Positive Sample:} 

Represent this product to match a user sequence:

Start Smart Kids Fun Silicone Swim Cap for Boys and Girls - Sharks \& Minnows.

\item \textbf{Negative Sample:}

Encode this product to fit a user sequence:

Gamo Rocket .177 Cal, 9.6 Grains, Ballistic Tip, 150ct.

\end{itemize}

\paragraph{Query2Item}
\begin{itemize}

\item \textbf{Query:}

Create an embedding of this user query to retrieve a relevant item:

floral napkins baby shower

\item \textbf{Positive Sample:} 

Utilize the product's title to form a representation that corresponds to a user query:

Vintage Floral Paper Napkins for Bridal Shower, Tea Party \& Luncheon (6.5 x 6.5 In, 100 Pack)

\item \textbf{Negative Sample:}

Represent the given product with a title to match a user query:

AIMEILI Nail Dipping Powder Nail Art Powder 6 Colors for Dipping Nail
\end{itemize}

\begin{table}[t]
\centering
\caption{Ablation experiments result of generative tasks on different model variants.}
\resizebox{\textwidth}{!}{%
\begin{tabular}{lcccccccccc}
\toprule
\multicolumn{1}{c}{\multirow{3}{*}{Model Variants}} & \multicolumn{3}{c}{General NLP} & \multicolumn{4}{c}{User Understanding} & \multicolumn{3}{c}{Item Understanding} \\
\cmidrule(lr){2-4} \cmidrule(lr){5-8} \cmidrule(lr){9-11}
 & QR↑ & AVE↑ & AG↑ & SR↑ & SA↑ & UP↑ & AP↑ & PRP↑ & PM↑ & IP↑ \\
 & (F1) & (F1) & (F1) & (R@1) & (F1) & (F1) & (F1) & (F1) & (F1) & (F1) \\
\midrule
RecFound & \underline{0.973} & \textbf{0.344} & \textbf{0.853} & \textbf{0.251} & \underline{0.533} & \textbf{0.895} & \textbf{0.830} & \underline{0.462} & \textbf{0.994} & \textbf{0.903} \\
w/o M.M. & \textbf{0.974} & \underline{0.341} & \textbf{0.853} & 0.234 & \textbf{0.545} & \textbf{0.895} & \underline{0.827} & \textbf{0.471} & \underline{0.988} & \underline{0.902} \\
w/o M.M. \& S2Sched & 0.965 & 0.283 & \underline{0.852} & \underline{0.245} & 0.499 & 0.885 & 0.816 & 0.356 & \underline{0.988} & 0.896 \\
w/o M.M. \& TMoLE & 0.966 & 0.264 & 0.847 & 0.221 & 0.527 & \underline{0.887} & 0.807 & 0.385 & 0.976 & 0.901 \\
w/o M.M. \& S2Sched \& TMoLE & \underline{0.973} & 0.316 & 0.839 & 0.242 & 0.501 & 0.874 & 0.814 & 0.426 & 0.967 & 0.889 \\
\bottomrule
\end{tabular}%
}
\label{tab:abgen}
\end{table}

\begin{table}[t]
\centering
\caption{Ablation experiments result of embedding tasks on different model variants. }
\resizebox{0.75\textwidth}{!}{%
\begin{tabular}{lccc}
\toprule
\multicolumn{1}{c}{\multirow{2}{*}{Model Variants}} & Item Emb↑ & Query Emb↑ & User Emb↑\\ 
 & (NDCG@20) & (MRR@20) & (MRR@20)\\ 
\midrule
RecFound & \textbf{0.823} & \underline{0.993} & \textbf{0.510} \\
w/o M.M. & 0.822 & \textbf{0.995} & \underline{0.508} \\
w/o M.M. \& S2Sched & 0.820 & \underline{0.992} & 0.474 \\
w/o M.M. \& TMoLE & \underline{0.822} & 0.991 & 0.465 \\
w/o M.M. \& S2Sched \& TMoLE & 0.820 & 0.985 & 0.439 \\
\bottomrule
\end{tabular}%
}
\label{tab:abemb}
\end{table}

\paragraph{Item2Item}

\begin{itemize}

\item \textbf{Query:}

Using the product's title and categories, produce an embedding to retrieve items that are alike:

title: Terra Men's Pilot, categories: Fashion

\item \textbf{Positive Sample:}

Using the attributes and details of an item, create an embedding to help find similar products:

title: Terra Men's Pilot, categories: Fashion

\item \textbf{Negative Sample:}

Based on an item's details and attributes, produce an embedding to identify similar products:

title: Crew Socks For Men Women Glamour Hot Pink Glitter Cotton Performance Cushion Sport Athletic With Moisture Wicking, categories: Fashion

\end{itemize}

\section{Ablation Study Details}
\label{app:ablation}

Table~\ref{tab:abgen} and Table~\ref{tab:abemb} present the detailed results of the ablation study on generative and embedding tasks, respectively, while Table~\ref{tab:ablation} summarizes the average performance across both branches. The trends observed in the detailed task-level metrics are largely consistent with the averaged results, with performance gradually declining as more components are removed.

\end{document}